\documentclass[preprint,3p]{elsarticle}
\usepackage{lineno}
\usepackage[numbers]{natbib}
\usepackage{stfloats}
\usepackage{amssymb}
\usepackage{graphicx}
\usepackage{hyperref}
\usepackage{siunitx}
\usepackage{lineno}
\usepackage{graphicx}
\usepackage{amsmath}
\usepackage{color}
\usepackage{bm}
\usepackage{amssymb}
\usepackage{mathrsfs}
\usepackage{amsbsy}
\usepackage{caption}
\usepackage {ulem}
\usepackage[thicklines]{cancel}
\usepackage{diagbox}
\usepackage{subfigure}
\newcommand{\meqref}[1]{Eq. (\ref{#1})}
\newcommand{\mpref}[1]{Fig. \ref{#1}}
\newcommand{\mtref}[1]{Tab. \ref{#1}}
\newcommand{\be}{\begin{equation}}
\newcommand{\ee}{\end{equation}}
\newcommand{\bea}{\begin{eqnarray}}
\newcommand{\eea}{\end{eqnarray}}

\usepackage{chngcntr}
\biboptions{sort&compress}

\begin{document}

\begin{frontmatter}

\title{Thermodynamic Stability Versus Chaos Bound Violation in D-dimensional RN Black Holes: Angular Momentum Effects and Phase Transitions}

\author[math,phys]{Yu-Qi Lei}

\author[phys]{Xian-Hui Ge\corref{mycorrespondingauthor}}
\ead{gexh@shu.edu.cn}
\cortext[mycorrespondingauthor]{Corresponding author}

\author[phys]{Surojit Dalui}

\affiliation[math]{organization={Department of Mathematics, Shanghai University},
             addressline={99 Shangda Road},
             city={Shanghai},
             postcode={200444},
             state={},
             country={China}}

\affiliation[phys]{organization={Department of Physics, Shanghai University},
             addressline={99 Shangda Road},
             city={Shanghai},
             postcode={200444},
             state={},
             country={China}}

\begin{abstract}
We compute the Lyapunov exponents for test particles orbiting in unstable circular trajectories around D-dimensional Reissner-Nordstr\"om (RN) black holes, scrutinizing instances of the chaos bound violation. Notably, we discover that an increase in particle angular momentum exacerbates the breach of the chaos bound. Our research centrally investigates the correlation between black hole thermodynamic phase transitions and the breaking of the chaos limit. Findings suggest that the chaos bound can only be transgressed within thermodynamically stable phases of black holes. Specifically, in the four-dimensional scenario, the critical point of the thermodynamic phase transition aligns with the threshold condition that delineates the onset of chaos bound violation. These outcomes underscore a deep-rooted link between the thermodynamic stability of black holes and the constraints imposed by the chaos bound on particle dynamics.

\end{abstract}

\end{frontmatter}


\section{Introduction}\label{introduction}
\noindent
 Black hole thermodynamics stands as a cornerstone in the broader landscape of general relativity, intertwining its threads with information theory, quantum attributes of black holes, and statistical physics principles\cite{Wald:1999vt,Page:2004xp}. Over the past several decades, substantial research efforts have delved into the multifaceted thermodynamics of black holes. A seminal contribution by Davies  \cite{Davies:1977bgr} revealed that Kerr and Reissner-Nordström (RN) black holes exhibit a discontinuous divergence in their heat capacity at constant charge or angular momentum, marking a critical phase transition point, colloquially referred to as the Davies point. This discovery significantly enriches our comprehension of the complex and nuanced properties of black holes. Recent investigations have turned their focus towards elucidating the intricate ties between the Lyapunov exponent governing particle motion and the thermodynamic phase structures of black holes\cite{Wei:2019jve,Lan:2020fmn,Guo:2022kio,Yang:2023hci}. Numerical evidence in \cite{Wei:2019jve} demonstrates a compelling correspondence between the quasinormal modes and the Davies point in black hole systems. This association gains further traction in \cite{Lan:2020fmn}, where quasinormal modes are shown to have a direct bearing on the phase transitions of regular black holes, with the Davies point serving as a central pivot. The intrigue deepens when the Lyapunov exponent enters the fray; works in  \cite{Guo:2022kio} and \cite{Yang:2023hci} convincingly establish a fundamental link between the phase transitions of black holes and their associated Lyapunov exponents.

 Moreover, the Lyapunov exponent of a black hole system operates under an established upper limit, famously known as the chaos bound \cite{Maldacena:2015waa}. Given this intimate connection between the Lyapunov exponent and black hole phase transitions, probing the chaos bound offers a novel lens through which to explore the intricacies of black hole thermodynamics and the dynamical behavior of particles in proximity to the event horizon. Hence, this paper embarks on a comprehensive examination of the thermodynamic stability of black holes and the implications of the chaos bound on particle motion, aiming to shed light on the fundamental interplay between these two seemingly disparate yet inherently intertwined aspects of black hole physics.

 The Lyapunov exponent in black hole systems is fundamentally constrained by a celebrated upper limit, recognized as the \textit{chaos bound} \cite{Maldacena:2015waa}, where $\lambda \leq \frac{2\pi T}{\hbar}$, indicative of the maximal degree of chaos in thermal quantum systems. Maldacena, Shenker, and Stanford expounded upon this through quantum field theory and via shock wave analyses near black hole horizons \cite{Maldacena:2015waa}. Hashimoto and Tanahashi subsequently generalized this notion to test particles in close proximity to the horizon, setting the bound as $\lambda \leq \kappa$, consistent with the thermodynamic identity $\kappa = 2\pi T$, thereby solidifying the harmony between the chaos bound in particle dynamics and quantum field theory results \cite{Hashimoto:2016dfz}. They exposed the inherent connection between the instability of particle motion within black holes and the chaos bound. Building on these findings, Dalui and Majhi extended the chaos bound to Kerr black holes \cite{Dalui:2018qqv} and Rindler horizons \cite{Dalui:2019umw}, yet this remains an evolving chapter in the narrative. Recent scholarly pursuits have scrutinized the chaos bound by examining static equilibrium states of test particles exterior to various charged black holes, discovering that under certain exceptional black hole configurations, the bound $\lambda \leq \kappa$ can be surpassed \cite{Zhao:2018wkl, Lei:2020clg}. Such breaches imply heightened instability in particle motion within spacetime. Further investigation has incorporated the influence of angular momentum, leading to extensive discussions on the chaos bound across a broad spectrum of black holes, including RN(-AdS) \cite{Lei:2021koj}, Kerr-Newman(-AdS/dS) \cite{Kan:2021blg, Gwak:2022xje, Park:2023lfc}, Einstein-Maxwell-Dilaton-Axion \cite{Yu:2022tlr}, and numerous others \cite{Gao:2022ybw, Chen:2023wph, Jeong:2023hom, Yu:2023spr, Xie:2023tjc}, wherein the angular momentum of particles presents additional avenues for chaos bound violation, primarily concentrating on the Lyapunov exponent associated with particles' unstable circular orbits.

The connection between chaos around black holes and gauge/gravity duality was first discussed by the chaotic dynamic of ring string \cite{PandoZayas:2010xpn}. Insights into the relationship between chaotic particle/string trajectories and the chaos bound can be found in \cite{Bera:2021lgw, Ma:2022tvs}. Meanwhile, noteworthy research has highlighted the profound physical implications of the Lyapunov exponent in particle motion, revealing its intimate ties to particle momentum and energy \cite{Hashimoto:2021afd}, causality \cite{Hashimoto:2022kfv}, minimal length effects\cite{Lu:2018mpr, Guo:2020pgq} and the null energy condition \cite{Giataganas:2021ghs, Lei:2023jqv}. Moreover, the interplay between near-horizon instability and quantum thermal properties of the black hole horizon has been systematically studied \cite{Dalui:2019esx, Dalui:2020qpt}.

  Despite these advancements, the complete physical meaning of the chaos bound and its violation in particle motion, a ubiquitous feature of black holes, remains elusive. It has frequently been observed that within specific parameter domains, the chaos bound can indeed be broken, yet the underlying rationale remains obscure. What is more, the exact physical mechanism facilitating this violation, the microscopic degrees of freedom accountable for it, and the subsequent repercussions are yet to be deciphered. These inquiries are crucial for advancing our comprehension of the fundamental unity between seemingly distinct yet fundamentally interconnected facets of black hole physics and chaos.

So, the points which are mentioned above are one of the grey areas of black hole physics which are not well explained for decades. In this work, for the first time, we have tried to give a possible reason based on our findings for such a feature. Here, we show a new perspective on the intrinsic connection between the thermodynamic stability and the violation of chaos bound $\lambda \leq \kappa$ by analyzing the heat capacity $C_Q$ of $D$ dimensional RN black holes. Here we focus on the asymptotically flat case, and refrain from discussing the AdS and dS cases. We discover that the Lyapunov exponent of unstable circular orbits always increases with the angular momentum of the test particles, thereby increasing the likelihood of exceeding the upper bound on the Lyapunov exponent. With the large angular momentum limit, we identify a threshold value $\bar{r}_c$ that indicates the conditions under which $\lambda \leq \kappa$ are violated. This parameter signifies that when the black hole's event horizon radius $r_+$ satisfies $r_+ > \bar{r}_c$, the condition $\lambda \leq \kappa$ is always met. By comparing the black hole's thermodynamic phase transition point $r_D$ with $\bar{r}_c$, we observe that violations of the chaos bound only occur in the phase of small black holes with thermodynamic stability. Therefore, the whole analysis establishes the fact that there must be some intriguing connection between the thermal stability of the black hole and the violation of the chaos bound and thereby demands a possible candidate to explore the underlying reason for this fascinating phenomenon.

The rest is organized as follows. In the next section, we provide a brief review of the thermodynamics of $D$ dimensional RN black holes. In Sec.\ref{Sec3}, we calculate the Lyapunov exponent of the unstable circular trajectory of test particles and discuss the relationship between the thermodynamic stability of the black hole and the violation of the chaos bound. Finally, we discuss and summarize in Sec.\ref{Sec4}.

\section{D dimensional Reissner-Nordstr\"om (RN) black hole and its thermodynamics}
\begin{figure*}[!ht]
\centering
\subfigure[$D=4$ and $r_D=\sqrt{3} \approx 1.732$.]{\label{cqd4}
\begin{minipage}[t]{0.3\linewidth}
\centering
\includegraphics[width=1 \textwidth]{ 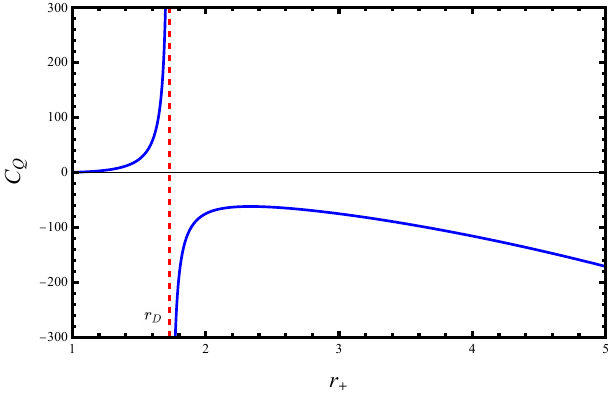}
\end{minipage}
}
\subfigure[$D=5$ and $r_D= \left(\frac{10}{3\pi} \right)^{1/4} \approx 1.015$.]{\label{cqd5}
\begin{minipage}[t]{0.3\linewidth}
\centering
\includegraphics[width=1 \textwidth]{ 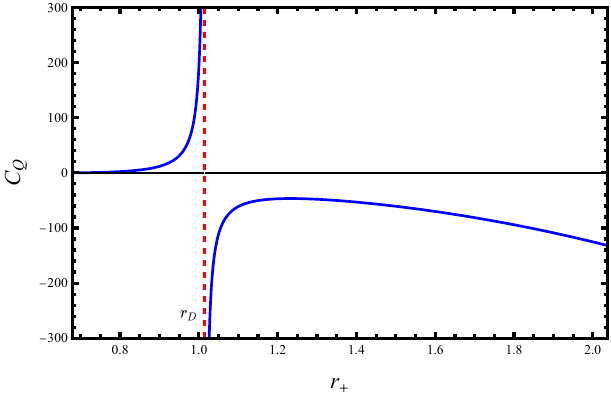}
\end{minipage}
}
\subfigure[$D=6$ and $r_D=\left(\frac{7}{4\pi} \right)^{1/6} \approx 0.907$.]{\label{cqd6}
\begin{minipage}[t]{0.3\linewidth}
\centering
\includegraphics[width=1 \textwidth]{ 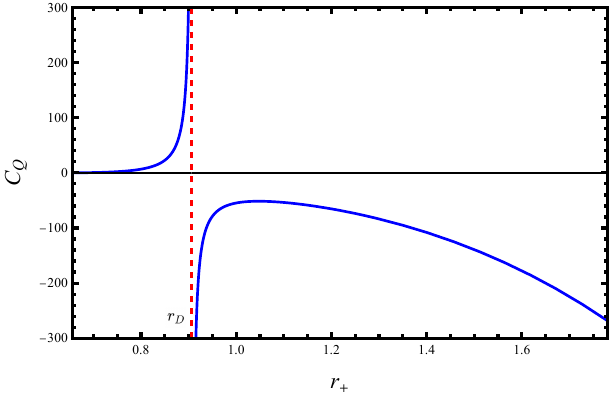}
\end{minipage}
}
\caption{Heat capacity $C_Q$ at constant charge $Q=1$ with $r_+$ for $D$ dimensional RN black holes. The figure (a), (b), (c) correspond to $D=4,\ 5,\ 6$, respectively.}
\label{CQ}
\end{figure*}

\noindent
Let us start with the $D$-dimensional RN black hole which can be reproduced from the charged Myers-Perry black hole \cite{Banerjee:2010ye,Pradhan:2013bli}. The black hole metric is 
\be
\begin{aligned}
ds^2=-f(r)dt^2+\frac{dr^2}{f(r)}+d\Omega_{D-2}^2, \quad f(r)=1-\frac{2 \Bar{M}}{r^{D-3}}+\frac{\Bar{Q}}{r^{2(D-3)}}.
\end{aligned}\label{metricf}
\ee
The largest root for $f(r)=0$ is the outer horizon $r_+$ of the black hole. The parameters $\Bar{M}$ and $\Bar{Q}$ are related to the Arnowitt-Deser-Misner(ADM) mass $M$ and charge $Q$ of black hole
\be
\begin{aligned}
M=&\frac{(D-2)\omega_{D-2} }{ 8 \pi} \Bar{M},
\\
Q^2=&\frac{(D-2)(D-3)\omega_{D-2}}{8\pi}  \Bar{Q}^2,
\end{aligned}
\ee
where $\omega_{D-2} = \frac{2\pi^{\frac{D-1}{2}}}{\Gamma(\frac{D-1}{2})}$ is the area of the unit sphere in $D-2$ dimensions. The corresponding electromagnetic potential $A_{t}$ satisfies
\be
A_t(r)=\frac{Q}{(D-3)r^{D-3}}.\label{funAt}
\ee
With $f(r_+)=0$, the black hole mass $M$ can be expressed as
\be 
M=\frac{Q^2 r_+^{3-D}}{2D-6}+\frac{(D-2)\pi^{\frac{D-3}{2}}r_+^{D-3}}{8\Gamma (\frac{D-1}{2})}.\label{funM}
\ee
With the Bekenstein-Hawking formula, we can obtain the black hole entropy $S$
\be
S=\frac{A_H}{4}=\frac{\omega_{D-2}}{4}r_+^{D-2},
\ee 
where $A_H$ is the area of horizon. The $D$ dimensional RN black hole satisfies the first law of thermodynamics
\be 
\delta M=T \delta S+\Phi \delta Q,
\ee 
where $T$ denotes the temperature of black hole, and $\Phi$ is the potential of the black hole. Based on this thermodynamic relationship, we can obtain the temperature $T$, the potential $\Phi$ and the heat capacity $C_Q$
\begin{small}
\be
\begin{aligned}
T=&\frac{1}{4\pi r_+}\left(D-3-\frac{4\pi^{\frac{1}{2}(3-D)}r_+^{6-2D}\Gamma(\frac{D-2}{2}) Q^2 }{D-2} \right),
\\
\Phi=&\frac{Q}{(D-3)r_+^{D-3}},
\\
C_Q=&\frac{2(D-2)^2\pi^{D+\frac{1}{2}}r_+^{3D}T}{4(2D-5)\pi^{\frac{3}{2}}Q^2r_+^{7}\Gamma^2(\frac{D-1}{2})-(D-2)(D-3)\pi^{\frac{D}{2}}r_+^{2D+1}\Gamma(\frac{D-1}{2})},
\end{aligned}\label{Thp}
\ee
\end{small}
where $C_Q$ is obtained at the constant $Q$. Therefore, the expression of the surface gravity $\kappa$ we obtain in this case is the following
\be 
\kappa=2 \pi T= \frac{1}{2 r_+}\left(D-3-\frac{4\pi^{\frac{1}{2}(3-D)}r_+^{6-2D}\Gamma(\frac{D-2}{2}) Q^2 }{D-2} \right).\label{funkappa}
\ee
To avoid the naked singularity, it is necessary to satisfy the inequality
\be
r_+ \geq r_e =2^{\frac{1}{2(D-3)}}\left(\frac{\pi^{\frac{1}{4}(3-D)}\sqrt{\Gamma (\frac{D-3}{2})}}{\sqrt{D-2}} \right)^{\frac{1}{D-3}},
\ee 
where $r_e$ is the radius of extremal black hole. For the non-extremal cases, there are some singularities in its heat capacity $C_Q$. From the expression of $C_Q$, we can obtain that $C_Q$ diverges when satisfying 
\be
4(2D-5)\pi^{\frac{3}{2}}Q^2r_+^{6}\Gamma (\frac{1}{2}(D-1))-(D-2)(D-3)\pi^{\frac{D}{2}}r_+^{2D}=0.
\ee
We mark the phase transition point as $r_+=r_D$, which is also called as Davies points. The phase transition point $r_D$ is 
\be
r_D=\left(\frac{\sqrt{D-2} \pi^{\frac{D+1}{4}}}{2\sqrt{(2D-5)Q^2\Gamma (\frac{D-3}{2})}}\right)^{\frac{1}{3-D}}
\ee
With the expression in \meqref{Thp} we plot the heat capacity at constant charge $C_Q$ as a function of the radius of horizon $r_+$ for $D$ dimensional RN black holes in \mpref{CQ}, here we take $D=4,\ 5,\ 6$ and the charge of black hole $Q=1$. The phase transition point $r_D$ is marked in the red dashed line. In the range of $r_+<r_D$, the heat capacity $C_Q$ is positive, which means that the black hole is thermodynamically stable till this point, whereas for $r_+>r_D$ the specific heat $C_Q$ is negative means from this point on the black hole is in the region of thermodynamically unstable phase.

\section{Thermodynamic stability and the violation of chaos bound}\label{Sec3}
\noindent
We next review the Lyapunov exponent of the particle's circular motion and explore the relationship between thermodynamic stability and the upper bound on the Lyapunov exponent. However, before jumping into that discussion let us first discuss briefly the Lyapunov exponent for a particle's circular motion and its connection to the angular momentum of that same test particle.

\subsection{The Lyapunov exponent}
\noindent
The motion of test particles near the black hole can be effectively represented by an inverse harmonic oscillator potential. At the maximum of the effective potential, the test particle has an unstable equilibrium in the radial direction (we consider here as the circular motion in $(r,\ \phi)$ space)
\be
\frac{\mathrm{d}^{2} r }{\mathrm{d} t^{2}}=\lambda ^2(r-r_0),
\ee
where $r_0$ is the radial position of circular orbits, and the parameter $\lambda$ is related to the form of the effective potential. The equation of motion corresponding to the circular motion has the general solution given by
\be
r=r_0+C_1 e^{\lambda t}+C_2 e^{-\lambda t},
\ee
where $C_1$ and $C_2$ are integration constants. This result shows that when the circular motion is perturbed by $\epsilon$, the perturbation $\epsilon$ grows exponentially as $\epsilon \sim e^{\lambda t}$, with the exponent $\lambda$ representing the Lyapunov exponent.

The above provides a concise overview of using the effective potential method to calculate the Lyapunov exponent for the circular orbit of a particle. By varying the charge, energy and angular momentum of test particles, we can find the position where the effective potential reaches its maximum value, thereby maintaining radial equilibrium for the particle. For specific calculation details, please refer to \cite{Hashimoto:2016dfz,Kan:2021blg,Gwak:2022xje}. In this paper, we will employ the Jacobian matrix method to calculate the Lyapunov exponent. Details of the calculation on the Lyapunov exponent for unstable circular orbits can be found in \ref{AppA}. The Lyapunov exponent $\lambda$ satisfies the following equation at the radial equilibrium position $r=r_0$
\begin{small}
\be
\lambda^2=\left.\frac{1}{4}\Bigg[f^{' 2}-\frac{4L^2f^{2}\left(A_t^{'}\left(2L^2+3r^2 \right) +rA_t^{''}\left(L^2+r^2 \right)\right)}{r^2\left(L^2+r^2 \right)^2A_t^{'}} +f\Bigg\{f^{'}\left(\frac{4L^2}{rL^2+r^3}+\frac{2A_t^{''}}{A_t^{'}} \right) \Bigg\}-2f^{''}\Bigg]\right|_{r=r_0}.\label{lambdaFull}
\ee
\end{small}
For convenience in discussion, we take the positive angular momentum of test particles ($L>0$).

To see the effect of the angular momentum $L$ of test particles, we can consider the near-horizon expansion, which means the position of the circular orbit of the particle motion is now situated very close to the horizon $r_+$. As detailed in \ref{AppB}, we assume the equilibrium position $r_0=r_++\delta$, where $\delta$ is a small positive quantity. Consequently, the expanded expression for $\lambda^2$ near the horizon $r_+$ can be rewritten as follows
\begin{eqnarray}
\lambda^2(r_{0}-r_{+})&&=\lambda^2\Big|_{r_{0}=r_{+}}+\frac{\partial}{\partial r_{0}}(\lambda^2)\Big|_{r_{0}=r_{+}}(r_{0}-r_{+})+\frac{\partial^{2}}{\partial r_{0}^{2}}(\lambda^{2})\Big|_{r_{0}=r_{+}}\frac{(r_{0}-r_{+})^{2}}{2!}+...\nonumber\\
&&=\kappa^2+\gamma_1 \delta +\gamma_2 \delta^2 +\mathcal{O}\left(\delta^3 \right),
\end{eqnarray}
where the first-order expansion parameter $\gamma_1$\footnote{We have written down the expression of the second-order expansion parameter $\gamma_2$ in \ref{AppB} and have discussed its effect there.}
\begin{small}
\be 
\gamma_1=4\kappa^2 \left(\frac{L^2}{L^2 r_++ r_+^3} +\frac{A_t^{''}(r_+)}{2 A_t^{'}(r_+)}\right).\label{gamma1}
\ee 
\end{small}
\noindent
From the expression of \(\gamma_1\), it is evident that as the angular momentum \(L\) increases, the value of \(\gamma_1\) also increases. This indicates that the Lyapunov exponent for the unstable circular orbit near the horizon is larger. Therefore, a higher \(L\) increases the likelihood of violating the chaos bound \(\lambda \leq \kappa\).

However, the relationship between angular momentum and chaos bound is not limited to the near-horizon region; it is a more general inference, and the results in some papers also support it \cite{Kan:2021blg,Lei:2021koj,Gwak:2022xje,Lei:2023jqv,Xie:2023tjc}. To understand the violation of chaos bound more, we can discuss the Lyapunov exponent of test particles in the limit of large angular momentum. Let us refer back to the expression of $\lambda^2$  in \meqref{lambdaFull}, which is applicable for any position outside the horizon and can be written down in this form also
\begin{small}
$$
\lambda^2=\left.\frac{1}{4}\Bigg[f^{' 2}-\frac{4f^{2}\left(A_t^{'}\left(2+3(r/L)^2 \right) +rA_t^{''}\left(1+(r/L)^2 \right)\right)}{r^2\left(1+(r/L)^2 \right)^2A_t^{'}} +f\Bigg\{f^{'}\left(\frac{4}{r+(r/L)^3}+\frac{2A_t^{''}}{A_t^{'}} \right) \Bigg\}-2f^{''}\Bigg]\right|_{r=r_0}$$
\end{small}
Now, in the large angular momentum limit (i.e. $(r/L)\rightarrow 0$), we can obtain the corresponding Lyapunov exponent $\bar{\lambda}$ by
\begin{small}
\be
\bar{\lambda}^2=\left.\frac{1}{4}\Bigg[f^{'2}-\frac{4f^2\left(2A_t^{'}+rA_t^{''} \right)}{r^2A_t^{'}} +f\Bigg\{
    f^{'} \left(\frac{4}{r}+\frac{2A_t^{''}}{A_t^{'}} \right)-2f^{''}\Bigg\}\Bigg]\right|_{r=r_0}.\label{lambdaLL}
\ee
\end{small}

\subsection{The numerical results of $\lambda^2-\kappa^2$}
\noindent
Next, we explore the relationship between the Lyapunov exponent of unstable circular orbits of test particles outside $D$-dimensional Reissner-Nordström (RN) black holes and the chaos bound $\lambda \leq \kappa$. To more clearly present the results, we use the sign of $\lambda^2-\kappa^2$ as an indicator of whether the chaos bound is violated. When $\lambda^2-\kappa^2>0$, it indicates that the chaos bound is violated. The values of $\lambda$ can be calculated using \meqref{lambdaFull}, and the scenario of the large angular momentum limit can be addressed by \meqref{lambdaLL}. Meanwhile, the surface gravity $\kappa$ is obtained from \meqref{funkappa}. The necessary metric function $f(r)$, electric potential function $A_t(r)$ and the related parameter relationships can be found in \meqref{metricf}-\ref{funM}.

\begin{figure*}[htb!]
\centering
\subfigure[$D=4$ and $L=1$.]{\label{d4pl1}
\begin{minipage}[t]{0.3\linewidth}
\centering
\includegraphics[width=1 \textwidth]{ 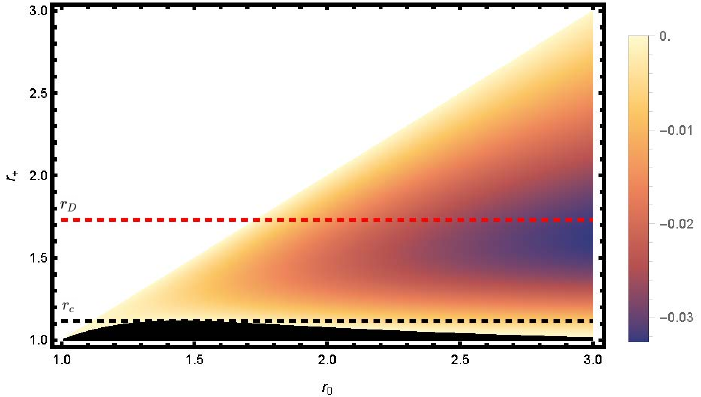}
\end{minipage}
}
\subfigure[$D=4$ and $L=5$.]{\label{d4pl5}
\begin{minipage}[t]{0.3\linewidth}
\centering
\includegraphics[width=1 \textwidth]{ 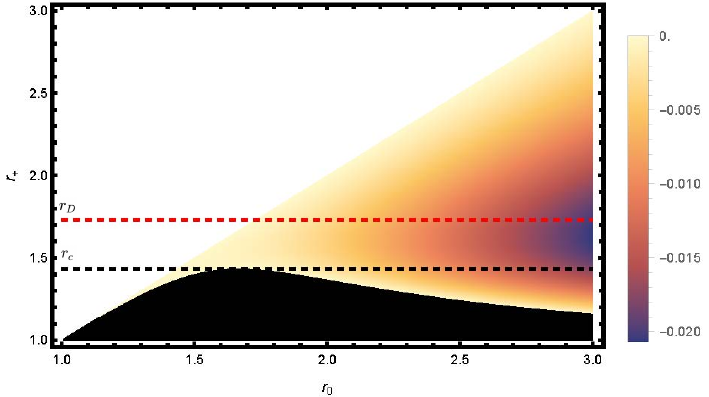}
\end{minipage}
}
\subfigure[$D=4$ and the large angular momentum limit.]{\label{d4pL}
\begin{minipage}[t]{0.3\linewidth}
\centering
\includegraphics[width=1 \textwidth]{ 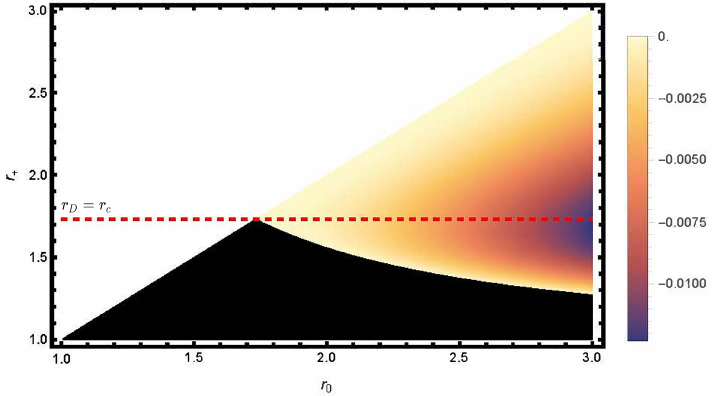}
\end{minipage}
}
\caption{The numerical results of $\lambda^2-\kappa^2$ in 4-dimensional RN black holes. Sub-figures (a) and (b) are calculated using \meqref{lambdaFull}, and sub-figure (c) uses \meqref{lambdaLL}.}
\label{d4pLy}
\end{figure*}

\begin{figure*}[htb!]
\centering
\subfigure[$D=5$ and $L=1$.]{\label{d5pl1}
\begin{minipage}[t]{0.3\linewidth}
\centering
\includegraphics[width=1 \textwidth]{ 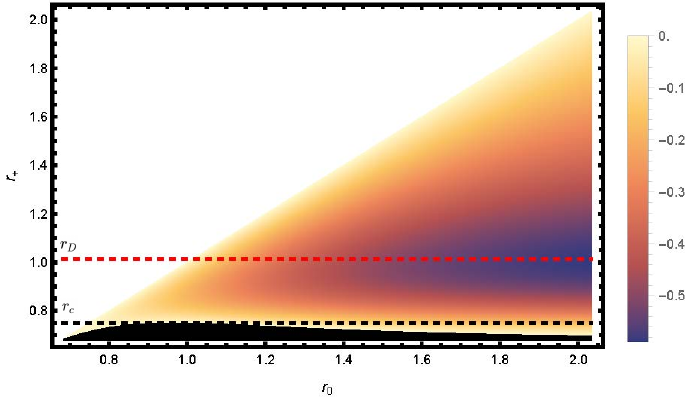}
\end{minipage}
}
\subfigure[$D=5$ and $L=5$.]{\label{d5pl5}
\begin{minipage}[t]{0.3\linewidth}
\centering
\includegraphics[width=1 \textwidth]{ 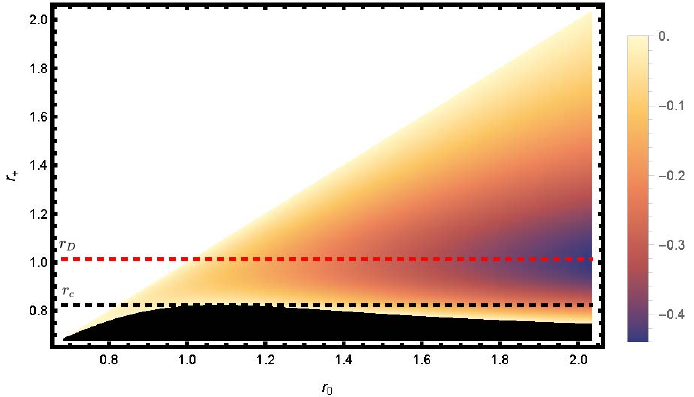}
\end{minipage}
}
\subfigure[$D=5$ and the large angular momentum limit.]{\label{d5pL}
\begin{minipage}[t]{0.3\linewidth}
\centering
\includegraphics[width=1 \textwidth]{ 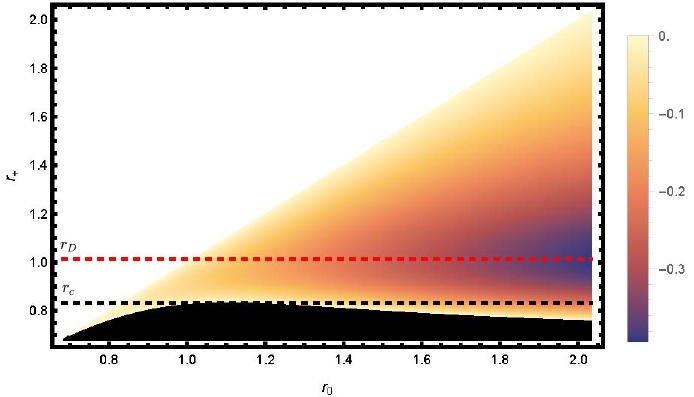}
\end{minipage}
}
\subfigure[$D=6$ and $L=1$.]{\label{d6pl1}
\begin{minipage}[t]{0.3\linewidth}
\centering
\includegraphics[width=1 \textwidth]{ 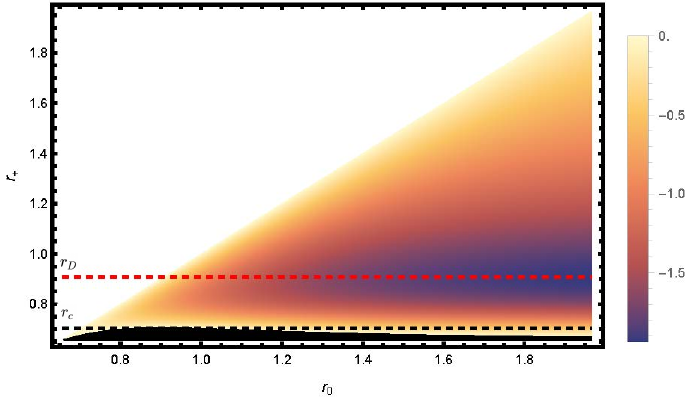}
\end{minipage}
}
\subfigure[$D=6$ and $L=5$.]{\label{d6pl5}
\begin{minipage}[t]{0.3\linewidth}
\centering
\includegraphics[width=1 \textwidth]{ 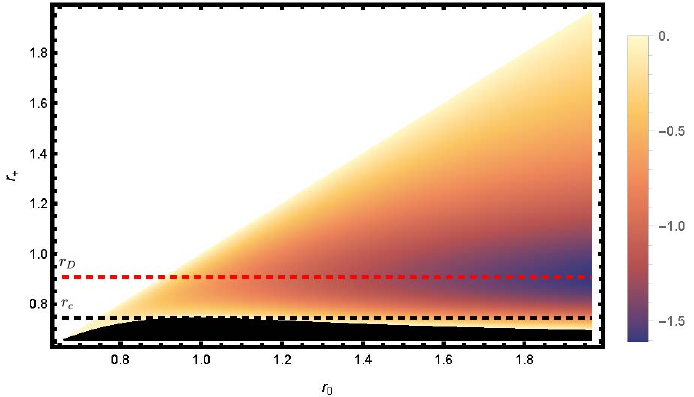}
\end{minipage}
}
\subfigure[$D=6$ and the large angular momentum limit.]{\label{d6pL}
\begin{minipage}[t]{0.3\linewidth}
\centering
\includegraphics[width=1 \textwidth]{ 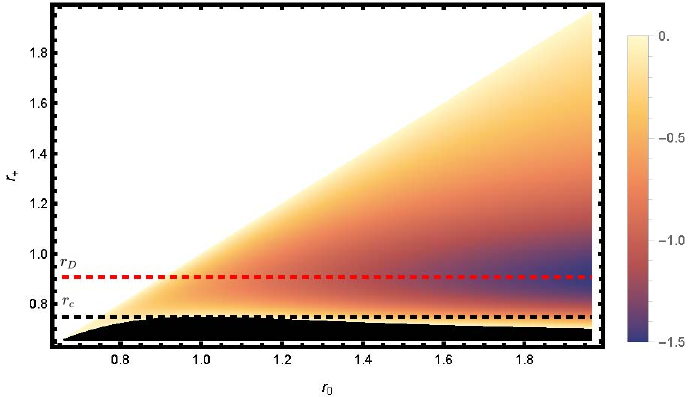}
\end{minipage}
}
\caption{The numerical results of $\lambda^2-\kappa^2$ in 5-dimensional and 6-dimensional RN black holes. Sub-figures (a), (b),(d) and (e) are calculated using \meqref{lambdaFull}, and sub-figure (c) and (f) use \meqref{lambdaLL}.}
\label{d56pLy}
\end{figure*}

In \mpref{d4pLy} and \mpref{d56pLy}, we plot the density map of the numerical results for $\lambda^2-\kappa^2$ on the $(r_0,r_+)$ plane, where the circular motion position $r_0$ and the black hole horizon radius $r_+$ both vary within the range of $(r_e,5r_e)$, with $r_e$ denoting the horizon radius corresponding to an extremal black hole. The influence of $r_+$ is incorporated as a parameter in the metric function $f(r)$, and the relationship is detailed in \meqref{metricf}-\ref{funM}. In these figures, we fix the black hole charge 
$$
Q=1.
$$
We focus on particle movements outside the black hole horizon, requiring that the circular orbit radius $r_0$ and the horizon radius $r_+$ satisfy $r_+/r_0 < 1$. Results for this scenario are shown in the bottom right part of each figure. Areas in white, indicating $r_+/r_0 > 1$, are not considered in our study.

To enhance the comprehensibility of the information presented in the figures, we use colored sections to denote results where $\lambda^2-\kappa^2<0$, indicating that the chaos bound is not exceeded; black sections correspond to areas where $\lambda^2-\kappa^2>0$, signifying that the bound $\lambda \leq \kappa$ is violated. Additionally, we draw the thermodynamic phase transition point of the black hole, $r_D$, with a red dashed line in the figures, and mark a threshold value $r_c$ with a black dashed line, which demonstrates that chaos bound violations no longer occur when $r_+>r_c$. By comparing the positions of $r_D$ (red dashed line) and $r_c$ (black dashed line) in the figures, we discuss the relationship between violations of the chaos bound and thermodynamic phase transition points in $D$-dimensional RN black holes. 

We present the results for 4-dimensional RN black holes \mpref{d4pLy}, where \mpref{d4pl1}, \mpref{d4pl5} and \mpref{d4pL} correspond to particle angular momentum values of $L=1$, $L=5$ and the large angular momentum limit, respectively. The Lyampunov exponents are calculated using \meqref{lambdaFull} and \meqref{lambdaLL}. Comparing these three figures, we observe that the range of the black regions, which indicate the violation of the chaos bound, expands as the particle's angular momentum increases. This implies that the particle's angular momentum indeed enhances the breaking of the chaos bound. Furthermore, the black regions only appear when $r_+<r_D$. In \mpref{d4pL}, we also see that the threshold value $r_c$, beyond which the chaos bound can be violated in the large angular momentum limit, coincides with $r_D$. This suggests a consistency between the thermodynamic phase transition behavior and the violation of the chaos bound by particle circular orbits in 4-dimensional RN black holes, hinting at a close connection between chaotic dynamics in particle motion and black hole thermodynamics.

In \mpref{d56pLy}, we display the results for 5-dimensional and 6-dimensional RN black holes using the previously mentioned expression for $\lambda$. The first row presents the 5-dimensional case, where \mpref{d5pl1}, \mpref{d5pl5} and \mpref{d5pL} respectively correspond to results of $L=1$, $L=5$, and the large angular momentum limit. The bottom row shows the 6-dimensional case. The figures show that, in higher-dimensional scenarios, the increase in particle angular momentum still leads to a stronger violation of the chaos bound. A notable difference, however, is that the consistency between the violation of the chaos bound at the large angular momentum limit and the black hole's thermodynamic phase transition no longer holds. Nonetheless, it is consistently observed that the threshold value $r_c$, associated with the chaos bound violation, is always less than the thermodynamic phase transition point $r_D$.

The values of threshold parameter \( r_c \) for the chaos bound violations discussed here can be found in \mtref{Trc}, which are obtained by the numerical analysis of the values of $\lambda^2-\kappa^2$. From the table, it can be seen that the value of \( r_c \) decreases as the spacetime dimension \( D \) increases. More importantly, \( r_c \) increases with the angular momentum \( L \), confirming our analysis by the near-horizon expansion that angular momentum increases the likelihood of violating the chaos bound. This result is not confined to the near-horizon region.

\begin{table}[htp!]
    \centering
    \begin{tabular}{|c|c|c|c|}
    \hline 
    \diagbox{Dimension}{Angular momentum $L$}&$L=1$&$L=5$&Large $L$ limit\\
    \hline
    D=4 & $r_c=1.116$ & $r_c=1.433$ &$r_c=1.732$  \\
    \hline
    D=5 & $r_c=0.750$ & $r_c=0.823$ & $r_c=0.833$ \\
    \hline
    D=6 & $r_c=0.703$ & $r_c=0.744$ & $r_c=0.748$\\
    \hline
    \end{tabular}
    \caption{The values of threshold parameter $r_c$ for violating the chaos bound in different cases.}\label{Trc}
\end{table}

\subsection{More discussions }
\noindent
Our numerical study on the upper bound of the Lyapunov exponent and its violations in $D$-dimensional RN black holes reveals that as the angular momentum of the particle increases, the violation of the chaos bound becomes more pronounced. We further analyze the relationship between the violation of the bound and the properties of black holes. To study the violation on $\lambda \leq \kappa$, we analyze the threshold value $r_c$ here. As mentioned previously, $r_c$ delineates the parameter range within which the chaos bound can be violated; for $r_+ > r_c$, the bound remains unviolated. 

\begin{figure*}[htp!]
\centering
\subfigure[$D=4$.]{\label{d4l}
\begin{minipage}[t]{0.3\linewidth}
\centering
\includegraphics[width=1 \textwidth]{ 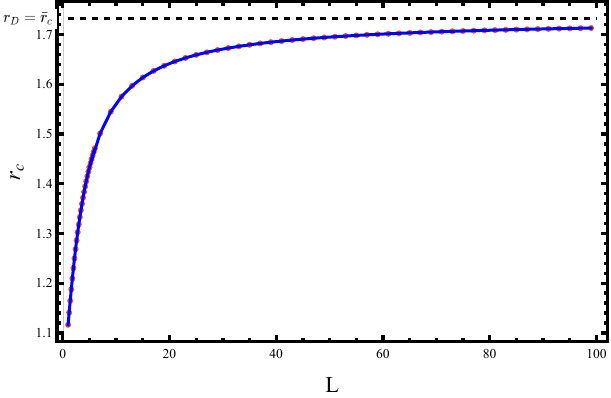}
\end{minipage}
}
\subfigure[$D=5$.]{\label{d5l}
\begin{minipage}[t]{0.3\linewidth}
\centering
\includegraphics[width=1 \textwidth]{ 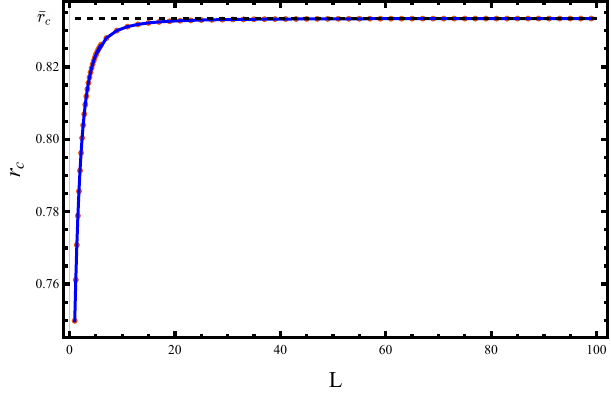}
\end{minipage}
}
\subfigure[$D=6$.]{\label{d6l}
\begin{minipage}[t]{0.3\linewidth}
\centering
\includegraphics[width=1 \textwidth]{ 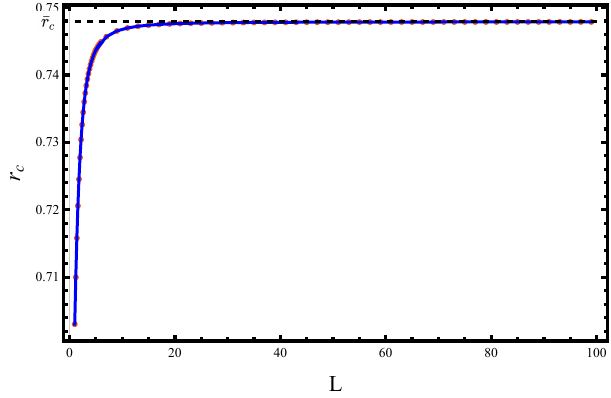}
\end{minipage}
}
\caption{The threshold value $r_c$ for violating chaos bound as a function of the angular momentum $L$ in $D$-dimensional RN black holes. The figures (a), (b), (c) correspond to $D=4,\ 5,\ 6$, respectively.}
\label{dl}
\end{figure*}

We can determine the threshold value $r_c$ for each angular momentum $L$ by analyzing the numerical results of $\lambda^2-\kappa^2$. We plot the threshold value $r_c$ as a function of the test particle's angular momentum $L$ in \mpref{dl}. The red dots represent results of $r_c$ for different angular momentum values, and the blue solid line is fit to these points. The topmost black dashed line in these plots represents the result for the large angular momentum limit, denoted by $\bar{r_c}$, and it also can be defined by $\bar{r}_c = \max(r_c)$. In the figures, it is observed that $r_c$ increases with $L$, approaching $\bar{r_c}$ in the case of large angular momentum. This result indicates that $\lambda > \kappa$ is only possible within the range $r_+ < \bar{r_c}$.

In our results about $D$-dimensional RN black holes ($D=4,\ 5,\ 6$), there is always $r_D \geq \bar{r}_c$, when $D=4$ it takes equal. That means the chaos bound only can be violated in the stable thermodynamic phase near the extremal black hole.

\section{Conclusions}\label{Sec4}
\noindent
In this work, we explore the relationship between the violation of the chaos bound in particle motion outside $D$-dimensional RN black holes and the thermodynamic stability of black holes. We calculated the Lyapunov exponents for unstable circular orbits of particles outside black holes and discussed the cases for $D=4, 5, 6$, showing that the violation of $\lambda \leq \kappa$ becomes increasingly apparent with the increase in the angular momentum of particles. In higher-dimensional cases, the chaos bound is violated within a smaller region. In the limit of large angular momentum, we derived the threshold value $\bar{r}_c$ from the numerical results of the Lyapunov exponent, indicating that the chaos bound is no longer broken when $r_+ > \bar{r}_c$. By comparing with the thermodynamic phase transition point $r_D$ of RN black holes, we find $\bar{r}_c \leq r_D$, suggesting that the violation of the chaos bound only occurs in the thermodynamically stable phase near extremal black holes with positive heat capacity.\footnote{In the 4-dimensional case, we have $\bar{r}_c = r_D$, which demonstrates the consistency between thermodynamic phase transitions and the violation of $\lambda \leq \kappa$ in 4-dimensional RN black holes.} This result reveals the possible intrinsic connection between the chaos bound in particle motion and the thermodynamic properties of black holes.

Previous studies have reported on the violation of the chaos bound in particle motion, yet the physical significance of this violation has remained elusive. The relationship between thermodynamic stability and dynamic stability warrants further in-depth investigation, especially in the context of black hole physics. Considering the recent findings that reveal a potential connection between the Lyapunov exponent and the temperature within thermodynamically stable regions of  D-dimensional RN black holes, the question arises whether such a correlation between the Lyapunov exponent and temperature holds a universal character across different systems. To establish this as a general principle, more empirical evidence and theoretical scrutiny are required to substantiate the existence of a persistent link between these two key variables in the broader scope of thermodynamic and dynamical stabilities. Although it is not a conclusive remark, we need to study it further. 

This finding merits further investigation, and it would be interesting to verify the universality of this conclusion across other black holes. For future work, we propose extending our analysis using the geometrothermodynamics (GTD) framework as discussed by \cite{Banerjee:2016nse}). Their metric-independent approach to phase transitions using geometrothermodynamics (GTD) offers us a potentially powerful tool for generalizing our conclusions. By considering the singularities of the curvature scalar in the GTD framework, we can gain deeper insights into the universal properties of phase transitions across different black hole spacetimes. By leveraging the GTD methodology, in the future, we aim to uncover universal patterns and deepen our understanding of chaos bound violations in black hole thermodynamics.

\newcounter{lastfigure}
\setcounter{lastfigure}{\value{figure}}

\section*{Acknowledgement} We would like to thank Bum-Hoon Lee for the helpful discussions. This work was partly supported by NSFC, China (Grant No. 12275166 and No. 12311540141). 

\appendix

\renewcommand{\thefigure}{\arabic{figure}}
\renewcommand{\thetable}{\arabic{table}}

\section{Lyapunov exponent}\label{AppA}
We consider test particles with mass $m$ and charge $e$ moving on the equatorial plane of black hole, and write the Lagrangian
\be 
\mathcal{L}=\frac{1}{2}\left(-f(r)\dot{t}^2+\frac{\dot{r}^2}{f(r)}+r^2\dot{\phi}^2 \right)-q A_t(r) \dot{t},
\ee
where $q=e/m$ and `` $\cdot$ '' denotes the derivative of the proper time $\tau$. With the definition of generalized momentum $\pi_\mu =\frac{\partial \mathcal{L}}{\partial \dot{x}^{\mu}}$, we can write the component
\be 
\begin{aligned}
\pi_t=&\frac{\partial \mathcal{L}}{\partial \dot{t}}=-f(r)\dot{t}-qA_t(r)=-E=\text{Constant},
\\
\pi_r=&\frac{\partial \mathcal{L}}{\partial \dot{r}}=\frac{\dot{r}}{f(r)},
\\
\pi_{\phi}=&\frac{\partial \mathcal{L}}{\partial \dot{\phi}}=r^2 \dot{\phi}=L=\text{Constant},
\end{aligned}
\ee 
where $E$ and $L$ represent the energy and angular momentum of test particles, respectively.

Using the relation $\mathcal{H}=\pi_\mu \dot{x}^\mu - \mathcal{L}$, we can obtain the Hamiltonian
\be 
\mathcal{H}=\frac{\pi_\phi^2f+r^2(\pi_r^2f^2-(\pi_t+qA_t)^2)}{2r^2f}.
\ee 
The equations of motion in proper time can be given by
\be 
\begin{aligned}
\dot{x}^{\mu}=\frac{\partial \mathcal{H}}{\partial \pi_\mu},\qquad \dot{\pi}_\mu=-\frac{\partial \mathcal{H}}{\partial x^{\mu}}.
\end{aligned}
\ee
Then let us write the radical motion in coordinate time $t$
\begin{small}
\be
\begin{aligned}
\frac{dr}{dt}=&\frac{\dot{r}}{\dot{t}}=\frac{\pi_r f^2}{E-q A_t},
\\
\frac{d\pi_r}{dt}=&\frac{\dot{\pi}_r}{\dot{t}}=\frac{f^{'}\left(qA_t-E\right)}{2f}+\frac{f\left(2L^2-r^3\pi_r^2f^{'}\right)}{2r^3\left(E-qA_t\right)}-qA_t^{'}.\label{eomt}
\end{aligned}
\ee 
\end{small}

Taking $(r,\pi_r)$ as the phase space variables, we can obtain the Jacobian matrix of particle motion, here we mark $\frac{dr}{dt}=F_1$ and $\frac{d\pi_r}{dt}=F_2$ for convenience
\be 
K_{ij}=
\left(
\begin{matrix}
\frac{\partial F_1}{\partial r}& \frac{\partial F_1}{\partial \pi_r}
\\
\frac{\partial F_2}{\partial r}& \frac{\partial F_2}{\partial \pi_r}
\end{matrix}
\right).
\ee 

For the circular motion at $r=r_0$, the Jacobian matrix can be reduced. We consider the circular motion condition $\pi_r=\frac{d\pi_r}{dt}=0$ and the 4-velocity normalization condition $\dot{x}^{\mu}\dot{x}_{\mu}=-1$, these conditions can be rewritten as 
\begin{small}
\be 
\begin{aligned}
q=&\left.\frac{2L^2f-rf^{'}\left(L^2+r^2 \right)}{2r^2A_t^{'}\sqrt{f\left(L^2+r^2 \right)}}\right|_{r=r_0},
\\
E=&\left.\frac{2rfA_t^{'}\left(L^2+r^2 \right) +A_t\left(2fL^2-rf^{'}\left(L^2+r^2 \right) \right) }{2r^2A_t^{'}\sqrt{f(L^2+r^2)}}\right|_{r=r_0}.
\end{aligned}\label{vecc}
\ee 
\end{small}
Then, the components of the Jacobian matrix are
\begin{small}
\be 
\begin{aligned}
K_{11}=&0,
\\
K_{12}=&\left.-\frac{f^2}{qA_t-E}\right|_{r=r_0},
\\
K_{21}=&\left.\left(\frac{(qA_t-E)f^{'}}{2f}\right)^{'} -qA_t^{''}-\left(\frac{L^2f}{r^3(qA_t-E)} \right)^{'}\right|_{r=r_0},
\\
K_{22}=&0.
\end{aligned}
\ee 
\end{small}

The Lyapunov exponent of test particles' circular orbits can be given by $\lambda^2=K_{12} K_{21}$ from the above functions.

\begin{small}
$$
\lambda^2=\left.\frac{1}{4}\left(f^{' 2}-\frac{4L^2f^{2}\left(A_t^{'}\left(2L^2+3r^2 \right) +rA_t^{''}\left(L^2+r^2 \right)\right)}{r^2\left(L^2+r^2 \right)^2A_t^{'}} +f\left(f^{'}\left(\frac{4L^2}{rL^2+r^3}+\frac{2A_t^{''}}{A_t^{'}} \right) \right)-2f^{''}\right)\right|_{r=r_0}
$$
\end{small}

\section{Near-horizon expansion and the higher-order effect}\label{AppB}
\noindent
To explore the Lyapunov exponent of particle motion near horizon, the near-horizon expansion is considered. Here we derive the expression for the Lyapunov exponent under the near-horizon expansion along with the corresponding expansion coefficients, and analyze the effect of higher-order terms. 

For the unstable circular motion of test particles near horizon, we can suppose its radial position $r_0=r_++\delta$ with $\delta >0$ and $\delta$ is a small value. Subsequently, by expanding \meqref{lambdaFull} to second order around $\delta\rightarrow0$ we obtain the expression for $\lambda^2$ under the near-horizon expansion
$$
\lambda^2=\kappa^2+\gamma_1 \delta +\gamma_2 \delta^2 +\mathcal{O}\left(\delta^3 \right),
$$
where the expansion parameters $\gamma_1$ and $\gamma_2$ are
\begin{small}
\be
\begin{aligned}
\gamma_1&=4\kappa^2 \left(\frac{L^2}{L^2 r_++ r_+^3} +\frac{A_t^{''}(r_+)}{2 A_t^{'}(r_+)}\right)
\\
\gamma_2&=\frac{1}{4}f^{'}(r_+)\left(\frac{6L^2\left(r_+ (L^2+r_+^2)f^{''}(r_+)-2(L^2+2r_+^2)f^{'}(r_+) \right)}{r_+^2(L^2+r_+^2)^2} -f^{(3}(r_+)\right)
\\
&-\frac{f^{'}(r_+)^2A_t^{''}(r_+)^2}{2A^{'}(r_+)^2}+\frac{3f^{'}(r_+)f^{''}(r_+)A_t^{''}(r_+)+2f^{'}(r_+)^2 \left(A_t^{(3)}(r_+)-\frac{2L^2 A_t^{''}(r_+)}{r_+L^2+r_+^3} \right) }{4A_t^{'}(r_+)}.
\end{aligned}\label{gamma2}
\ee
\end{small}

In our previous work on the 4-dimensional case, we discussed how \(\gamma_1\) approaches zero with increasing angular momentum \cite{Lei:2021koj}. Therefore, it becomes necessary to consider the impact of the second-order expansion coefficient \(\gamma_2\) the limit of large angular momentum. Then we obtained the threshold value for the violation of the chaos bound in 4-dimensional RN black holes as \(\bar{r}_c = \sqrt{3}Q\).

\begin{figure}[!ht]
\centering
\subfigure[$D=5$, the first-order expansion coefficient $\gamma_1$.]{\label{d5g1}
\begin{minipage}[t]{0.42\linewidth}
\centering
\includegraphics[width=1 \textwidth]{ 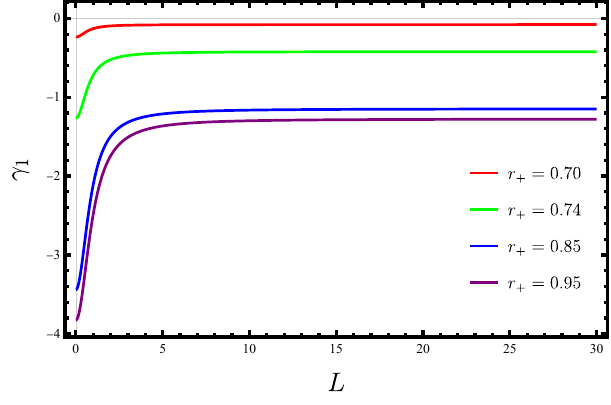}
\end{minipage}
}
\subfigure[$D=5$, the second-order expansion coefficient $\gamma_2$.]{\label{d5g2}
\begin{minipage}[t]{0.42\linewidth}
\centering
\includegraphics[width=1 \textwidth]{ 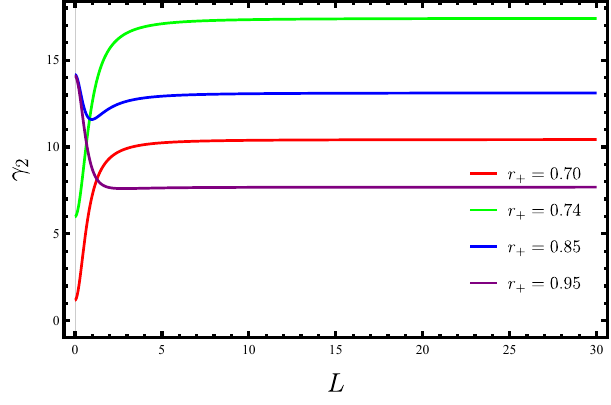}
\end{minipage}
}
\subfigure[$D=6$, the first-order expansion coefficient $\gamma_1$.]{\label{d6g1}
\begin{minipage}[t]{0.42\linewidth}
\centering
\includegraphics[width=1 \textwidth]{ 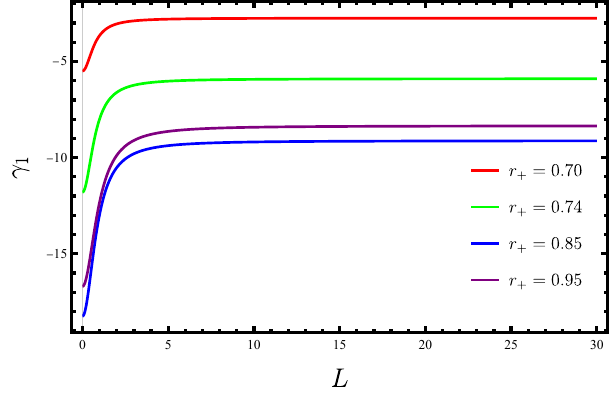}
\end{minipage}
}
\subfigure[$D=6$, the second-order expansion coefficient $\gamma_2$.]{\label{d6g2}
\begin{minipage}[t]{0.42\linewidth}
\centering
\includegraphics[width=1 \textwidth]{ 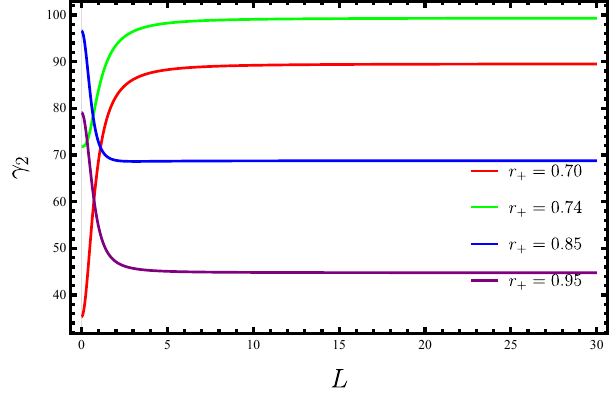}
\end{minipage}
}
\caption{The near-horizon expansion coefficients $\gamma_1$ and $\gamma_2$ as the function of particle angular momentum 
$L$ for 5-dimensional and 6-dimensional RN black holes with the black hole charge $Q=1$. }\label{gammafig}
\end{figure}

Here we study the near-horizon expansion behavior in $5$ and $6$ dimensional cases. We plot the expansion coefficients $\gamma_1$ and $\gamma_2$ as the function of angular momentum $L$ in \mpref{gammafig}, and the coefficients can be obtained by \meqref{gamma2}. In these figures, we set the radius of horizon $r_+=0.70,\ 0.74,\ 0.85,\ 0.95$ with the unit black hole charge $Q=1$, the different results are plotted in red, green, blue and purple, respectively. In particular, the red and green lines represent examples where the chaos bound can be violated, while the blue and purple lines correspond to cases without the violation of the bound. In \mpref{d5g1} and \mpref{d6g1}, we observe that \(\gamma_1\) is always negative and increases with angular momentum, ultimately approaching a negative value. This indicates that the first-order term does not lead to the violation of \(\lambda \leq \kappa\). For the second-order term, we see in \mpref{d5g2} and \mpref{d6g2} that \(\gamma_2\) is positive, offering the possibility to exceed the Lyapunov exponent upper bound. With \(r_+=0.70, 0.74\), \(\gamma_2\) increases with \(L\), while for \(r_+=0.85, 0.95\), it seems decreases with the increase of $L$. It's observable that for the red and green lines, \(|\gamma_1|\) always trends towards smaller values (\(\gamma_1<0\)), and their corresponding \(\gamma_2\) tends towards larger values. This can explain why, in their associated black hole parameter spaces, the chaos bound can be violated. Therefore, as previous work and this work have shown, higher-order terms in the near-horizon expansion can provide the possibility of exceeding the Lyapunov exponent upper bound. Moreover, further research has revealed that the behavior of violating the chaos bound is not limited solely to the near-horizon region.


\end{document}